\documentclass[12pt]{article}
\usepackage{amsfonts,amsmath,amssymb,amsthm,esint}
\usepackage{scalerel}[2016/12/29]
\usepackage{epsfig,graphicx}
\usepackage{tikz}
\usepackage{float}
\usepackage{braket}
\usepackage{url,hyperref}
\usepackage{wasysym}
\usetikzlibrary{decorations,arrows}
\usetikzlibrary{decorations.markings}
\usetikzlibrary{shapes.geometric}
\usetikzlibrary{patterns}

\def\be{\begin{equation}}
\def\ee{\end{equation}}
\def\bea{\begin{eqnarray}}
\def\eea{\end{eqnarray}}
\def\tr{{\rm tr}\,}
\def\Tr{{\rm \bf Tr}\,}

\def\rme{\mathrm{e}}
\def\rmi{\mathrm{i}}

\def\Q{\mathrm{Q}}
\def\rme{\mathrm{e}}
\def\rmi{\mathrm{i}}

\definecolor{R}{RGB}{100,0,0}
\definecolor{G}{RGB}{0,100,0}
\definecolor{B}{RGB}{0,0,100}

\begin{document}

\title{On the algebraic area of cubic lattice walks}

\author{Li Gan$^*$}

\maketitle


\begin{abstract}
We obtain an explicit formula to enumerate closed random walks on a cubic lattice with a specified length and 3D algebraic area. The 3D algebraic area is defined as the sum of algebraic areas obtained from the walk's projection onto the three Cartesian planes. This enumeration formula can be mapped onto the cluster coefficients of three types of particles that obey quantum exclusion statistics with statistical parameters $g=1$, $g=1$, and $g=2$, respectively, subject to the constraint that the numbers of $g=1$ (fermions) exclusion particles of two types are equal.
\end{abstract}

* LPTMS, CNRS, Universit\'e Paris-Saclay, 91405 Orsay Cedex, France\\\indent {\it ~~li.gan92@gmail.com}

\section{Introduction}
The algebraic area in two dimensions is defined as the area swept by planar closed random walks, weighted by the winding number in each winding sector. The area is considered positive if the walk moves around the sector in a counterclockwise direction. In the continuous case, the probability distribution of the algebraic area $A$ enclosed by closed Brownian curves after a time $t$ is given by L\'evy's stochastic area formula (also known as L\'evy's law) \cite{Levy}
\be \label{Levy 2D} P(A)=\frac{\pi}{2t} \frac{1}{\cosh^2(\pi A/t)}.\ee
In the discrete case, a series of explicit algebraic area enumeration formulae \cite{Shuang, papers,honeycomb} for closed random walks on various lattices have recently been obtained from the Kreft coefficients \cite{Kreft} encoding the Schr\"odinger equation of quantum Hofstadter-like models \cite{Hofstadter} that describe a charged particle hopping on planar lattices coupled to a perpendicular magnetic field. Essentially the enumeration amounts to calculating the trace of the power of Hofstadter-like Hamiltonian and has an interpretation in terms of statistical mechanics of particles that obey exclusion statistics with an integer exclusion parameter $g$ ($g=0$ for bosons, $g=1$ for fermions, $g\geq 2$ for stronger exclusion than fermions). Figure 1 shows three examples of 2D lattice random walks: the square lattice walk corresponds to $g=2$ exclusion, the Kreweras-like chiral walk on a triangular lattice corresponds to $g=3$ exclusion, and the honeycomb lattice walk corresponds to a mixture of $g=1$ and $g=2$ exclusion, with an appropriate spectrum. Note that in the context of Hofstadter-like model, the algebraic area can be expressed as $\frac{1}{2}\oint ({\bf r} \times d{\bf r})\cdot{\bf B}$, where ${\bf B}=(0,0,1)$ and the integral is along the closed walk in the $xy$-plane.

\begin{figure}[H]
\begin{center} 
\includegraphics{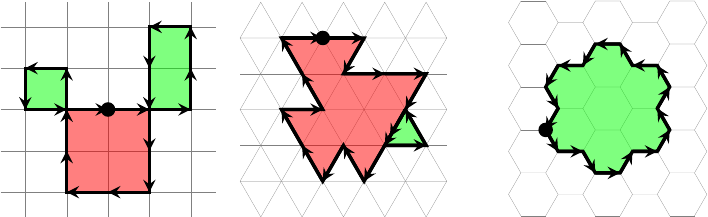}
\caption{\small Closed square lattice walk, chiral triangular lattice walk, and honeycomb lattice walk of length $18$ with algebraic area $-1$, $-14$ and $7$, respectively. The region inside the walk, i.e., winding sector, is colored green if its area is positive, otherwise it is colored red. In the chiral triangular lattice walk, only three of the possible six directions are allowed at each step.}\label{fig rw}
\end{center}
\end{figure}

\begin{figure}[H]
\begin{center}
\includegraphics{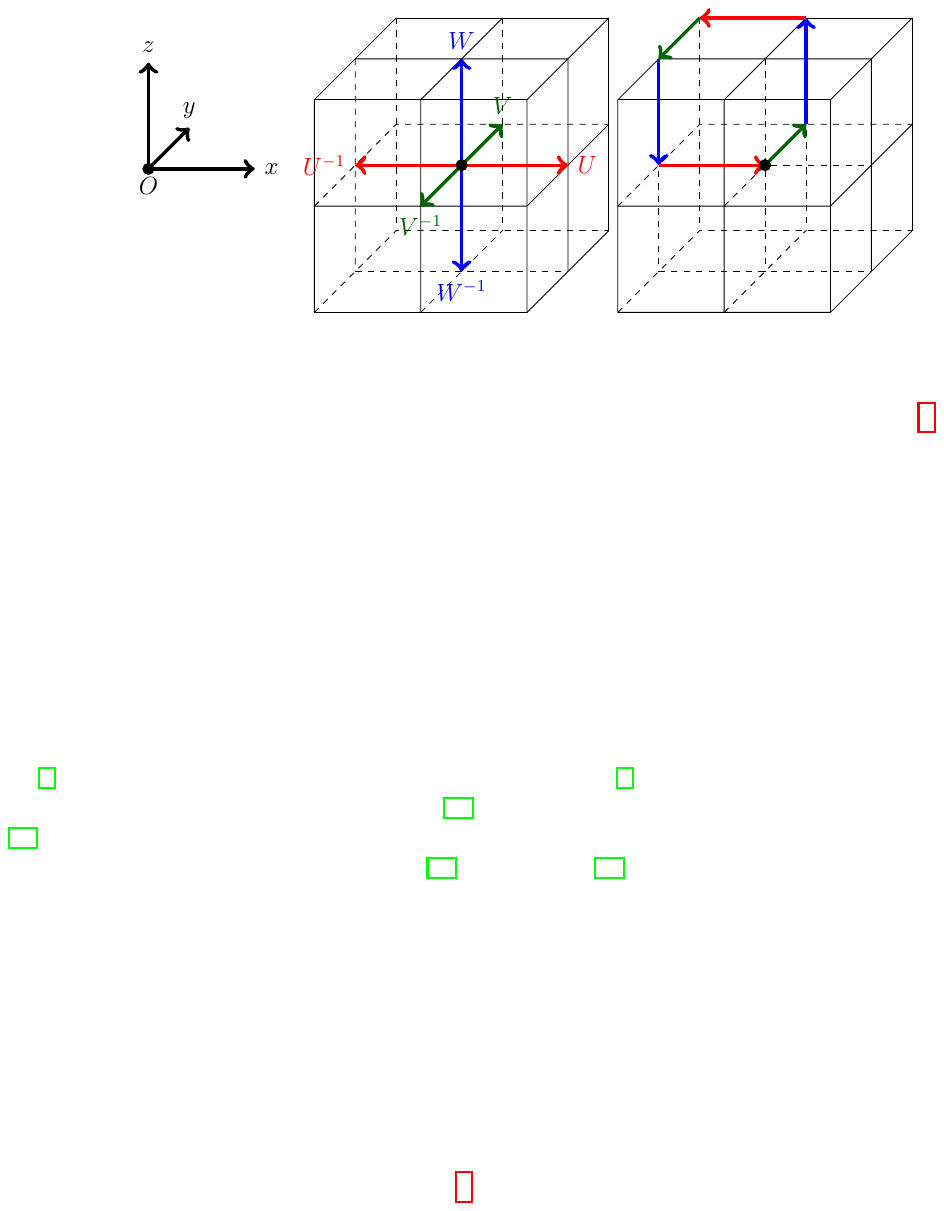}
\caption{\small(left) 3D Cartesian coordinate system; (middle) six lattice hopping operators in a cubic lattice; (right) closed 6-step cubic lattice walk $U W^{-1} V^{-1} U^{-1} W V$, whose 3D algebraic area is given by $A=1+1+(-1)=1$. Using the commutation relations (\ref{torus}), $U W^{-1} V^{-1} U^{-1} W V$ is simplified to $\Q^{1}$ as expected.}\label{fig cubic}
\end{center}
\end{figure}

In this article we extend the concept of algebraic area to closed 3D walks by defining it as the sum of the algebraic areas of the walk projected onto the $xy,yz,zx$-planes along the $-z,-x,-y$ directions. To count closed random walks on a cubic lattice with a given length and 3D algebraic area, we begin by introducing three lattice hopping operators $U, V, W$ along the $x,y,z$ directions, as well as $U^{-1},V^{-1},W^{-1}$ along the $-x,-y,-z$ directions. These operators satisfy the noncommutative 3-tori algebra \cite{Bedos}
\be \label{torus}
V \, U=\Q \, U \, V,
~W \, V=\Q \, V \, W,
~U \, W=\Q \, W \, U,
\ee
which is simply an alternative description of walks that go around the unit lattice cell on the Cartesian planes, i.e., $V^{-1}U^{-1}VU=\Q$, $W^{-1}V^{-1}WV=\Q$, and $U^{-1}W^{-1}UW=\Q$. The 3D algebraic area $A$ enclosed by a walk can thus be computed by reducing the corresponding hopping operators to $\Q^A$ using the commutation relations (\ref{torus}). See figure \ref{fig cubic} for the closed 6-step cubic lattice walk $U W^{-1} V^{-1} U^{-1} W V=\Q$ as an example. Another example involves enumerating closed 4-step walks. By taking the trace of $(U+V+W+U^{-1}+V^{-1}+W^{-1})^4=6(11+2\Q+2\Q^{-1})+\ldots$,
only terms with an equal number of $U$ and $ U^{-1}$, $V$ and $V^{-1}$, $W$ and $W^{-1}$ survive, yielding the count of algebraic area: 66 walks enclose an algebraic area $A = 0$, 12 walks enclose an algebraic area $A=1$, and 12 walks enclose an algebraic area $A=-1$.

By expressing the phase $\Q=\exp{(2\pi {\rmi} \phi / \phi_0)}$ in terms of the flux $\phi$ through the unit lattice cell on each of the three Cartesian planes in unit of the flux quantum $\phi_0$, the Hermitian operator
\be \label{cube Hamiltonian}
H=U+V+W+U^{-1}+V^{-1}+W^{-1}
\ee
represents a Hamiltonian that describes a charged particle hopping on a cubic lattice coupled to a magnetic field ${\bf B}=(1,1,1)$, as indicated in the definition of the 3D algebraic area. The energy spectrum with ${\bf B}=(1,1,1)$ on a cubic lattice was initially investigated in \cite{Hasegawa}. The 3D Hofstadter model was studied earlier in \cite{Montambaux}, and the general case of the uniform magnetic field was explored in \cite{Koshino}, with an experimental scheme proposed in \cite{Zhang}. Hofstadter models have also been studied on other 3D lattices, such as the tetragonal monoatomic and double-atomic lattice \cite{Bruning}, and in 4D \cite{Di Colandrea} as well.

As with the case of planar lattice, the quantum trace of $H^{2n}$ provides the generating function for the number $C_{2n}(A)$ of closed random walks of length $2n$ (necessarily even) on a cubic lattice enclosing a 3D algebraic area $A$. Specifically,
\be \label{C2n Tr}
\sum_{A}C_{2n}(A)\Q^A=\Tr H^{2n}
\ee
with the normalization $\Tr I=1$, where $I$ denotes the identity operator.

The paper is organized as follows: Assuming that the flux is rational, we use the finite-dimensional representation of the algebra (\ref{torus}) to derive the trace of $H^{2n}$, establish its connection with quantum exclusion statistics ($g=1$, $g=1$, $g=2$), and provide a combinatorial interpretation based on the combinatorial coefficients $c_{1,1,2}(\tilde{l}_1,\ldots,\tilde{l}_{j+1}; \tilde{l}'_1,\ldots,\tilde{l}'_{j+1}; l_1,\ldots,l_j)$ labeled by the $(1,1,2)$-compositions. In the Conclusion, we present the explicit formula for $C_{2n}(A)$, as well as its asymptotics  as $n\to \infty$, and discuss potential generalizations.

\section{Algebraic area enumeration of cubic lattice walks}
\subsection{Hamiltonian}
From now on, we assume that the magnetic flux on each Cartesian plane is rational, i.e., $\phi/\phi_0=p/q$ with $p$ and $q$ being coprime, thus $\Q=\exp{(2\pi{\rmi}p/q)}$. To obtain the finite-dimensional representation of $U,V,W$, we introduce the $q\times q$ ``clock" and ``shift" matrices
\be \nonumber \hspace{-1em}
u= {\rme}^{{\rmi}k_x} \begin{pmatrix}
\Q & 0 & 0 & \cdots & 0 & 0 \\
0 & \Q^{2}  &0& \cdots & 0 & 0 \\
0 & 0 & \Q^{3}  & \cdots & 0 & 0 \\
\vdots & \vdots & \vdots & \ddots & \vdots & \vdots \\
0 & 0 & 0  & \cdots &\Q^{q-1}  & 0 \\
0 & 0 & 0  & \cdots & 0 & 1\\
\end{pmatrix}, ~
v = {\rme}^{{\rmi}k_y} \begin{pmatrix}
\;0\; & \;1\; & \;0\; & \cdots & \;0\; & \;0\; \\
0 & 0 & 1 & \cdots & 0 & 0 \\
0 & 0 & 0 & \cdots & 0 & 0\\
\vdots & \vdots & \vdots &\ddots & \vdots & \vdots\\
0 & 0 & 0 & \cdots &0& 1\\
1 & 0 & 0 & \cdots & 0& 0 \\
\end{pmatrix},
\ee
which satisfy $v \, u=\Q \, u \, v$ and contribute to the Hofstadter Hamiltonian $u+v+u^{-1}+v^{-1}$ for square lattice walks. Here, $k_x$ and $k_y$ denote the quasimomenta in the $x$ and $y$ directions. In the quantum trace, integration over $k_x$ and $k_y$ eliminates the unwanted terms containing $u^q$ and $v^q$ which correspond to open walks but can be closed by $q$-periodicity. Another way to achieve this is by setting $k_x=k_y=0$ and considering walks of length less than $q$.

Because of the open walk $UVW\neq I$, it is not possible to represent the operators $U,V,W$ as $u,v,v^{-1}u^{-1}$, respectively, even though they satisfy the algebra (\ref{torus}). To address this, we introduce an additional vector space with dimension $q'$, in which $U$ and $V$ act as identity operators, while $W$ does not. Consequently, we obtain the representation of (\ref{torus}) as $q q' \times q q'$ matrices
$$U=u \otimes I_{q'},~V= v \otimes I_{q'},~W=(v^{-1}u^{-1}) \otimes u',$$
where $u'$ is an arbitrary $q' \times q'$ matrix. 
Again, the quasimomenta are set to be zero.
The sought-after quantum trace of the $qq' \times qq'$ Hamiltonian matrix (\ref{cube Hamiltonian}) reduces to the usual trace up to a normalization factor, that is,
$$\Tr H^{2n}=\frac{1}{q q'}\tr H^{2n}.$$
 
Let $u'$ be diagonal and equal to $u|_{q\to q'}$ (therefore $\Q\to \Q'=\exp(2\pi{\rmi}p/q')$ in $u'$). Performing the algebra-preserving transformation $u\to -u^{-1}v,~v\to v^{-1},~ u' \to -u'$ leads to the new Hamiltonian that describes walks on a deformed cubic lattice
$$H^{\prime}=H_2\otimes I_{q'} + u\otimes u' + u^{-1}\otimes u'^{-1},$$
where the Hofstadter Hamiltonian associated to the usual square lattice walks is
{\small \be \nonumber
H_2=-u^{-1}v-v^{-1}u+v+v^{-1}=
\begin{pmatrix}
0 & \bar{f}_1 & 0 & \cdots & 0 & 0 \\
f_1 & 0 & \bar{f}_2 & \cdots & 0 & 0 \\
0 & f_2 & 0 & \cdots & 0 & 0 \\
\vdots & \vdots & \vdots & \ddots & \vdots & \vdots \\
0 & 0 & 0 & \cdots & 0  & \bar{f}_{q-1} \\
~0~ & ~0~ & ~0~ & \cdots & f_{q-1} & ~0~ \\
\end{pmatrix}
\ee}with $f_k=1-\Q^{k}$. Note that $H_2$ is a $g=2$ matrix in the sense that its secular determinant $\det(I_q-zH_2)=\sum_{n=0}^{\lfloor q/2\rfloor}(-1)^n Z_n z^{2n}$ captures the Kreft coefficient \cite{Kreft}
$$Z_n= \sum_{k_1=1}^{q-2n+1} \sum_{k_2=1}^{k_1} \cdots \sum_{k_{n}=1}^{k_{n-1}}s_{k_1+2n-2}s_{k_2+2n-4} \cdots s_{k_{n-1}+2}s_{k_{n}},~Z_0=1$$
as a trigonometric multiple nested sum with $+2$ shifts among the spectral functions $s_{k}:=f_k \bar{f}_k =4\sin^2(k \pi p/q)$. In statistical mechanics, $Z_n$ can be interpreted as the $n$-body partition function for $n$ particles in a one-body spectrum $\epsilon_k~(k=1,2,\ldots,q-1)$ with Boltzmann factor ${\rme}^{-\beta \epsilon_k}=s_k$. The $+2$ shifts indicate that these particles obey $g=2$ exclusion statistics, i.e., no two particles can occupy two adjacent quantum states. By introducing cluster coefficients $b_n$ via $\log\Big(\sum_{n=0}^{\lfloor q/2 \rfloor}Z_n x^n\Big)=\sum_{n=1}^{\infty}b_n x^n$ with fugacity $x=-z^2$, and using the identity $\log \det(I_q-z H_2)=\tr \log(I_q-z H_2)=-\sum_{n=1}^{\infty}\frac{z^n}{n}\tr H^{n}_{2}$ we establish a connection between the generating function for algebraic area enumeration of square lattice walks and the cluster coefficients with $g=2$ exclusion statistics, that is,
\bea \nonumber \Tr H_2^{2n}&=&\frac{1}{q}\tr H_2^{2n}=2n(-1)^{n+1}\frac{1}{q}b_n \\
&=& 2n \hspace{-1em}
\sum_{\substack{l_1,l_2,\ldots,l_j\\\text{composition of }n}}
\hspace{-1em}
c_2(l_1,l_2,\ldots,l_j) \frac{1}{q}\sum_{k=1}^{q-j} s_k^{l_1} s_{k+1}^{l_2} \cdots s_{k+j-1}^{l_j}, \label{tr H2}
\eea
where $c_2(l_1,l_2,\ldots,l_j)=\frac{1}{l_1}\prod_{i=2}^{j}\binom{l_{i-1}+l_i-1}{l_i}$.
As we will see in Section \ref{112section}, the algebraic area enumeration for cubic lattice walks can also be mapped onto cluster coefficients with appropriate exclusion parameters and spectral functions.

Now come back to the Hamiltonian $H'$, where the matrix elements read
$$(H^{\prime})_{ij}=\tilde{s}_{\lceil i/q' \rceil, 1+ (i-1)\text{mod }q'}\delta_{ij}
+ f_{\lceil j/q' \rceil} \, \delta_{i, j+q'}
+ \bar{f}_{\lceil i/q' \rceil} \, \delta_{i, j-q'}
,~~i,j=1,2,\ldots,q q'$$
with $\tilde{s}_{k,k'}= \Q^k \Q'^{k'} + \Q^{-k} \Q'^{-k'}$.
Applying the trace computation techniques described in \cite{Dyck} we obtain
\bea \nonumber
\frac{1}{q q'}\tr H^{\prime 2n}
&=&
2n \hspace{-2em}
\sum_{\substack{\tilde{l}_1,\ldots,\tilde{l}_{j+1};l_1,\ldots,l_j\\(1,2){\text{-composition of }}2n}} \hspace{-2em}
c_{1,2}(\tilde{l}_1,\ldots,\tilde{l}_{j+1};l_1,\ldots,l_j)
\frac{1}{q}\sum_{k=1}^{q-j}
{s}_{k}^{{l}_1}
{s}_{k+1}^{{l}_2}
\ldots
{s}_{k+j-1}^{{l}_{j}}
\frac{1}{q'}
\sum_{k'=1}^{q'}
\tilde{s}_{k,k'}^{\tilde{l}_1}
\tilde{s}_{k+1,k'}^{\tilde{l}_2}
\cdots
\tilde{s}_{k+j,k'}^{\tilde{l}_{j+1}} \\\label{trace12}
& & 
\eea
\noindent with
\be \nonumber
c_{1,2}(\tilde{l}_1,\ldots,\tilde{l}_{j+1}; l_1,\ldots,l_j) =
\frac{(\tilde{l}_1+l_1-1)!}{\tilde{l}_1!~l_1!}
\prod_{k=2}^{j+1}\binom{l_{k-1}+\tilde{l}_k+l_k-1}{l_{k-1}-1,~\tilde{l}_k,~l_k}.
\ee
By convention $l_{k}=0$ for $k>j$. We define the sequence of integers $\tilde{l}_1,\ldots,\tilde{l}_{j+1};l_1,\ldots,l_j, j\geq 1$, as a $(1,2)$-composition of $2n$ if they satisfy the conditions
$$2n=(\tilde{l}_1 + \ldots + \tilde{l}_{j+1}) + 2(l_1+\ldots+l_j),~\tilde{l}_i \geq 0,~l_i>0,$$
i.e., $l_i$'s are the usual compositions of $1,2,\ldots,n$ and $\tilde{l}_i$'s are additional non-negative integers. For $j=0$ we have the trivial composition $\tilde{l}_1=2n$.

As $q'$ is arbitrary, for simplicity of calculation, we set $q'=q$ in the sequel. The second trigonometric sum in (\ref{trace12}) is expanded to be
\bea \nonumber
& & \frac{1}{q}\sum_{k'=1}^{q}
\tilde{s}_{k,k'}^{\tilde{l}_1}
\tilde{s}_{k+1,k'}^{\tilde{l}_2}
\cdots
\tilde{s}_{k+j,k'}^{\tilde{l}_{j+1}} \\\nonumber
&=& \frac{1}{q}
\sum_{\tilde{l}'_i+\tilde{l}''_i=\tilde{l}_i}
\Q^{(\tilde{l}'_2-\tilde{l}''_2)+2(\tilde{l}'_3-\tilde{l}''_3)+\ldots+j(\tilde{l}'_{j+1}-\tilde{l}''_{j+1})+k[(\tilde{l}'_1+\ldots+\tilde{l}'_{j+1})-(\tilde{l}''_1+\ldots+\tilde{l}''_{j+1})]} 
\sum_{k'=1}^{q}
\Q^{k'[(\tilde{l}'_1+\ldots+\tilde{l}'_{j+1})-(\tilde{l}''_1+\ldots+\tilde{l}''_{j+1})]} \\\nonumber
& & \times 
\binom{\tilde{l}_1}{\tilde{l}'_1}
\binom{\tilde{l}_2}{\tilde{l}'_2}
\cdots
\binom{\tilde{l}_{j+1}}{\tilde{l}'_{j+1}}
\eea
with $\tilde{l}'_i,\tilde{l}''_i \geq 0,~i=1,\ldots,j+1$.
Since $\sum_{k'=1}^{q}
\Q^{k'[(\tilde{l}'_1+\ldots+\tilde{l}'_{j+1})-(\tilde{l}''_1+\ldots+\tilde{l}''_{j+1})]}$ is non vanishing only when $\tilde{l}'_1+\ldots+\tilde{l}'_{j+1}=\tilde{l}''_1+\ldots+\tilde{l}''_{j+1}$ we obtain
$$\frac{1}{q}\sum_{k'=1}^{q}
\tilde{s}_{k,k'}^{\tilde{l}_1}
\tilde{s}_{k+1,k'}^{\tilde{l}_2}
\cdots
\tilde{s}_{k+j,k'}^{\tilde{l}_{j+1}}=
\sum_{\tilde{l}'_i+\tilde{l}''_i=\tilde{l}_i}
\Q^{
(\tilde{l}'_2-\tilde{l}''_2)+
2(\tilde{l}'_3-\tilde{l}''_3)+\ldots+
j(\tilde{l}'_{j+1}-\tilde{l}''_{j+1})
}
\binom{\tilde{l}_1}{\tilde{l}'_1}
\binom{\tilde{l}_2}{\tilde{l}'_2}
\cdots
\binom{\tilde{l}_{j+1}}{\tilde{l}'_{j+1}}.
$$
Finally, by recognizing that the binomial product $\displaystyle\binom{\tilde{l}_1}{\tilde{l}'_1}
\binom{\tilde{l}_2}{\tilde{l}'_2}
\cdots
\binom{\tilde{l}_{j+1}}{\tilde{l}'_{j+1}}$ can be absorbed into $c_{1,2}$, as well as changing the notation $\tilde{l}'_i \to \tilde{l}_i,~\tilde{l}''_i \to \tilde{l}'_i$, we arrive at
{\bea \nonumber
\frac{1}{q^2}\tr H^{\prime 2n}&=& 2n
{\hspace{-0.5em}} \sum_{\substack{\tilde{l}_1,\ldots,\tilde{l}_{j+1};\tilde{l}'_1,\ldots,\tilde{l}'_{j+1};l_1,\ldots,l_j \\ (1,1,2){\text{-composition of }}2n\\ \tilde{l}_1+\ldots+\tilde{l}_{j+1}=\tilde{l}'_1+\ldots+\tilde{l}'_{j+1}}} {\hspace{-0.5em}}
c_{1,1,2}(\tilde{l}_1,\ldots,\tilde{l}_{j+1}; \tilde{l}'_1,\ldots,\tilde{l}'_{j+1}; l_1,\ldots,l_j) \\ \label{tr}
& & \times
\Q^{
(\tilde{l}_2-\tilde{l}'_2)+
2(\tilde{l}_3-\tilde{l}'_3)+\ldots+
j(\tilde{l}_{j+1}-\tilde{l}'_{j+1})
}
\frac{1}{q}\sum_{k=1}^{q-j}
s_k^{l_1} s_{k+1}^{l_2} \ldots s_{k+j-1}^{l_j}
\eea}
\noindent with new combinatorial coefficients
\bea \nonumber
c_{1,1,2}(\tilde{l}_1,\ldots,\tilde{l}_{j+1}; \tilde{l}'_1,\ldots,\tilde{l}'_{j+1}; l_1,\ldots,l_j) &=&
\frac{(\tilde{l}_1+\tilde{l}'_1+l_1-1)!}{\tilde{l}_1!~\tilde{l}'_1!~l_1!}\prod_{k=2}^{j+1}\binom{l_{k-1}+\tilde{l}_k+\tilde{l}'_k+l_k-1}{l_{k-1}-1,~\tilde{l}_k,~\tilde{l}'_k,~l_k}.
\eea
By convention $l_{k}=0$ for $k>j$. We define the sequence of integers $\tilde{l}_1,\ldots,\tilde{l}_{j+1}; \tilde{l}'_1,\ldots,\tilde{l}'_{j+1}; l_1,\ldots,l_j$ as a $(1,1,2)$-composition of $2n$ if they satisfy the conditions
$$2n=(\tilde{l}_1 + \ldots + \tilde{l}_{j+1})+(\tilde{l}'_1+\ldots+\tilde{l}'_{j+1}) + 2(l_1+\ldots+l_j),
~\tilde{l}_i, \tilde{l}'_i \geq 0,~l_i>0,$$
i.e., $l_i$'s are the usual compositions of $1,2,\ldots,n$ and $\tilde{l}_i, \tilde{l}'_i$'s are non-negative integers. We also include, with constraint $\tilde{l}_1=\tilde{l}'_1$, the trivial composition $(n;n;0)$. A combinatorial interpretation of the $(1,1,2)$-composition and $c_{1,1,2}$ will be discussed in Section \ref{section com}.

\subsection{$(1,1,2)$-exclusion statistics}\label{112section}
Now we take a step further by defining $t_k=\Q^k$. Given that for $\tilde{l}_1+\ldots+\tilde{l}_{j+1}=\tilde{l}'_1+\ldots+\tilde{l}'_{j+1}$
\be \nonumber
\frac{1}{q}\sum_{k=1}^{q-j}
t_k^{\tilde{l}_1}\bar{t}_k^{\tilde{l}'_1}s_k^{l_1}
t_{k+1}^{\tilde{l}_2}\bar{t}_{k+1}^{\tilde{l}'_2}s_{k+1}^{l_2}\ldots=
\Q^{
(\tilde{l}_2-\tilde{l}'_2)+
2(\tilde{l}_3-\tilde{l}'_3)+\ldots+
j(\tilde{l}_{j+1}-\tilde{l}'_{j+1})
}
\frac{1}{q}\sum_{k=1}^{q-j}
s_k^{l_1} s_{k+1}^{l_2} \ldots s_{k+j-1}^{l_j},
\ee
we rewrite (\ref{tr}) in its standard form that consists solely of  compositions, combinatorial coefficient, and trigonometric sum, as follows:
\be \label{trace112} 
\frac{1}{q^2}\tr H^{\prime 2n}= 2n
{\hspace{-0.5em}} \sum_{\substack{\tilde{l}_1,\ldots,\tilde{l}_{j+1};\tilde{l}'_1,\ldots,\tilde{l}'_{j+1};l_1,\ldots,l_j \\ (1,1,2){\text{-composition of }}2n\\ \tilde{l}_1+\ldots+\tilde{l}_{j+1}=\tilde{l}'_1+\ldots+\tilde{l}'_{j+1}}} {\hspace{-0.5em}}
c_{1,1,2}(\tilde{l}_1,\ldots,\tilde{l}_{j+1}; \tilde{l}'_1,\ldots,\tilde{l}'_{j+1}; l_1,\ldots,l_j)
\frac{1}{q}\sum_{k=1}^{q-j}
t_k^{\tilde{l}_1}\bar{t}_k^{\tilde{l}'_1}s_k^{l_1}
t_{k+1}^{\tilde{l}_2}\bar{t}_{k+1}^{\tilde{l}'_2}s_{k+1}^{l_2}
\ldots,
\ee
which indicates a mixture of $g=1$, $g=1$, and $g=2$ exclusion. We call it $(1,1,2)$-exclusion statistics. Therefore,
\be \label{Tr}
\Tr H^{2n}=\frac{1}{q^2}\tr H^{2n}=\frac{1}{q^2}\tr H^{\prime 2n}=-\frac{2n}{q}b'_{2n}.\ee
That is, $\Tr H^{2n}$ is equivalent, up to a trivial factor, to the cluster coefficient $b'_{2n}$ associated with the $2n$-body partition function for particles in a one-body spectrum $\epsilon_k~(k=1,\ldots,q)$ obeying a mixture of three statistics: fermions with Boltzmann factor ${\rme}^{-\beta \epsilon_k}=t_k$, fermions of another type with Boltzmann factor ${\rme}^{-\beta \epsilon_k}=\bar{t}_k$, and two-fermion bound states occupying one-body energy levels $k$ and $k+1$ with Boltzmann factor ${\rme}^{-\beta \epsilon_{k,k+1}}=-s_k$ behaving effectively as $g=2$ exclusion particles. $b'_{2n}$ is {\it{constrained}} by the requirement that the numbers of the two types of fermions are equal, implying $\Tr H^{2n+1}=0$ as expected. Note that setting $t_k=\bar{t}_k=0$ in (\ref{trace112}) eliminates all terms with nonzero $\tilde{l}_i, \tilde{l}'_i$'s and (\ref{Tr}) effectively reduces to (\ref{tr H2}).
 
\subsection{Combinatorial interpretation}\label{section com}

\begin{figure}[H]
\begin{center}
\includegraphics{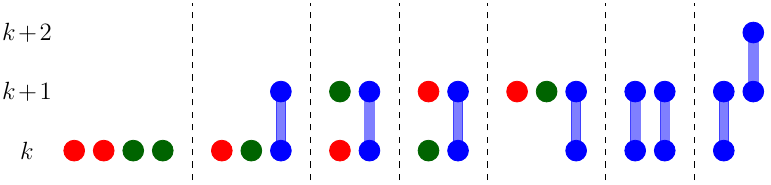}
\end{center}
\caption{\small{Seven $(1,1,2)$-compositions of $4$ with $\tilde{l}_1+\ldots+\tilde{l}_{j+1}=\tilde{l}'_1+\ldots+\tilde{l}'_{j+1}$: $(2;2;0)$, $(1,0;1,0;1)$, $(1,0;0,1;1)$, $(0,1;1,0;1)$, $(0,1;0,1;1)$, $(0,0;0,0;2)$, $(0,0,0;0,0,0;1,1)$, illustrated by two types of fermions (red, green) and two-fermion bound states (blue).}}\label{tr112}
\end{figure}

The $(1,1,2)$-compositions with the constraint $\tilde{l}_1+\ldots+\tilde{l}_{j+1}=\tilde{l}'_1+\ldots+\tilde{l}'_{j+1}$ have a combinatorial interpretation, which can be derived from their relation to cluster coefficients of $(1,1,2)$-exclusion statistics. Specifically, $(1,1,2)$-compositions of $2n$ with constraint correspond to all distinct connected arrangements of $2n$ particles on a one-body spectrum, consisting of two types of fermions (with equal numbers) and two-fermion bound states. In other words, they represent all the possible ways to place two types of particles and bound states on the spectrum such that they cannot be separated into two or more mutually non-overlapping groups. For example, as shown in figure \ref{tr112}, there are seven $(1,1,2)$-compositions of $4$ with $\tilde{l}_1+\ldots+\tilde{l}_{j+1}=\tilde{l}'_1+\ldots+\tilde{l}'_{j+1}$, which contribute to
{\small 
$$-b'_4=\frac{1}{4q}\tr H^{\prime 4} =
\frac{3}{2}\sum_{k=1}^{q}t_k^2 \bar{t}_k^2 + 
\sum_{k=1}^{q-1} \Big(
2 t_k \bar{t}_k +
 t_{k} \bar{t}_{k+1} + 
 t_{k+1}\bar{t}_k + 
2 t_{k+1} \bar{t}_{k+1} \Big) s_k +
\frac{1}{2}\sum_{k=1}^{q-1}s_k^2 +
\sum_{k=1}^{q-2}s_k s_{k+1}.$$}

\begin{figure}[H]
\begin{center}
\includegraphics{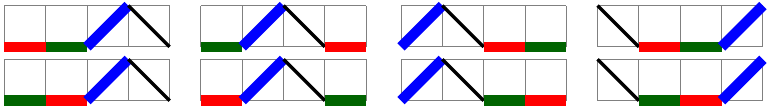}
\caption{\small All the $4c_{1,1,2}(1,0;1,0;1)=8$ periodic generalized Motzkin paths of length $2n=4$ with $\tilde{l}_1=1$ horizontal step (red),  $\tilde{l}'_1=1$ horizontal step of another type (green), and $l_1=1$ up step (blue) originating from the first floor.}\label{Motzkin path}
\end{center}
\end{figure}

Following the argument in \cite{Dyck}, $2n \, c_{1,1,2}(\tilde{l}_1,\ldots,\tilde{l}_{j+1}; \tilde{l}'_1,\ldots,\tilde{l}'_{j+1}; l_1,\ldots,l_j)$ admits an interpretation as the number of periodic generalized Motzkin paths of length $2n$ with $\tilde{l}_i$ horizontal steps, $\tilde{l}'_i$ horizontal steps of another type, and $l_i$ up steps originating from the $i$th floor (see figure \ref{Motzkin path} for an example).

\tikzset{hyper thick/.style={line width=5pt}}

\section{Conclusion}
Based on (\ref{C2n Tr}), (\ref{tr}), (\ref{Tr}) and the fact that the trigonometric sum $\frac{1}{q}\sum_{k=1}^{q-j} s_k^{l_1} s_{k+1}^{l_2} \ldots s_{k+j-1}^{l_j} $ can be computed \cite{Shuang,papers}, we deduce the desired counting for closed random walks on a cubic lattice with given length $2n$ and 3D algebraic area $A$
{\small \bea \nonumber \hspace{0em}
C_{2n}(A) &=& 2n
{\hspace{-0.6em}} \sum_{\substack{\tilde{l}_1,\ldots,\tilde{l}_{j+1};\tilde{l}'_1,\ldots,\tilde{l}'_{j+1};l_1,\ldots,l_j \\ (1,1,2){\text{-composition of }}2n\\ \tilde{l}_1+\ldots+\tilde{l}_{j+1}=\tilde{l}'_1+\ldots+\tilde{l}'_{j+1}}} {\hspace{-0.7em}}
\frac{(\tilde{l}_1+\tilde{l}'_1+l_1-1)!}{\tilde{l}_1!~\tilde{l}'_1!~l_1!}\prod_{k=2}^{j+1}\binom{l_{k-1}+\tilde{l}_k+\tilde{l}'_k+l_k-1}{l_{k-1}-1,~\tilde{l}_k,~\tilde{l}'_k,~l_k}
\sum_{k_3=-l_3}^{l_3} \sum_{k_4=-l_4}^{l_4}\cdots \sum_{k_j=-l_j}^{l_j} \\\nonumber
& & \binom{2l_1}{l_1+A-\sum_{i=2}^{j+1}(i-1)(\tilde{l}_i-\tilde{l}'_i)+\sum_{i=3}^{j}(i-2)k_i}
\binom{2l_2}{l_2-A+\sum_{i=2}^{j+1}(i-1)(\tilde{l}_i-\tilde{l}'_i)-\sum_{i=3}^{j}(i-1)k_i} \\\label{final formula}
& & \times \prod_{i=3}^{j}\binom{2l_i}{l_i+k_i}.
\eea}Note that the enumeration can be computed recursively as well. See Appendix A for further details and several examples of $C_{2n}(A)$. In Appendix B, we present some combinatorial results for $(1,1,2)$-compositions and $c_{1,1,2}$, where the overall counting of closed $2n$-step cubic lattice walks is recovered to be $\binom{2n}{n}\sum_{k=0}^{n}\binom{n}{k}^2 \binom{2k}{k}$.

In the continuum limit, in which the lattice spacing $a
\to 0$, closed cubic lattice walks become 3D Brownian motion loops. The probability distribution of the enclosing 3D algebraic area $A$ for a Brownian loop after a time $t$ is given by
\be \label{Levy 3D}
P'(A)=\frac{\pi}{2\sqrt{3}t}\frac{1}{\cosh^2[\pi A/(\sqrt{3}t)]}.
\ee
Note that this distribution is simply the Fourier transform of the partition function of a charged particle moving in continuous 3D space under a uniform magnetic field ${\bf B}=(1,1,1)$. By aligning the magnetic field with the $z$ direction through a change of coordinates, we obtain the standard Landau levels plus free motion in the $z$ direction. This explains why (\ref{Levy 3D}) coincides with L\'evy's law (\ref{Levy 2D}) for 2D Brownian loops, up to a rescaling of $A$ due to the normalization of ${\bf B}$. With the scaling ${\bf n}a^2=3t$, we infer from (\ref{Levy 3D}) the asymptotics for (\ref{final formula}) as the walk length ${\bf n}=2n\to\infty$
\be \label{asym} 
C_{\bf n}(A)
\sim
\frac{\sqrt{3}\pi}{2{\bf n}\cosh^2(\sqrt{3}\pi A / {\bf n})}{\binom{\bf n}{{\bf n}/2} \sum_{k=0}^{{\bf n}/2} \binom{{\bf n}/2}{k}^2 \binom{2k}{k}},\ee
where $A=0,\pm 1,\pm 2,\ldots$ is dimensionless. The asymptotics (\ref{asym}) has been checked numerically for ${\bf n}$ up to 42. However, deriving (\ref{asym}) directly from (\ref{final formula}) is nontrivial and remains an open problem.

It is natural to extend the definition of the 3D algebraic area to the sum of projection areas with \textit{arbitrary} weights, which is equivalent to specifying an arbitrary magnetic field ${\bf B}$. For instance, when ${\bf B}=(0,0,1)$, the 3D algebraic area is defined as the area of the walk projected onto the $xy$-plane. The counting for closed ${\bf n}$-step cubic lattice walks enclosing a 3D algebraic area $A$ under this definition turns out to be
$$C'_{\bf n}(A) = \sum_{l=0}^{{\bf n}/2} \binom{\bf n}{2l,{\bf n}/2-l,{\bf n}/2-l} C_{2l,{\text{sq}}}(A),$$
where $C_{2l,{\text{sq}}}(A)$ is the number of closed ${2l}$-step square lattice walks enclosing an algebraic area $A$. Similarly, as ${\bf n}\to \infty$,
$$C'_{\bf n}(A)
\sim
\frac{3\pi}{2{\bf n}\cosh^2(3\pi A / {\bf n})}{\binom{\bf n}{{\bf n}/2} \sum_{k=0}^{{\bf n}/2} \binom{{\bf n}/2}{k}^2 \binom{2k}{k}}.$$

\begin{figure}[H]
\begin{center}
\includegraphics{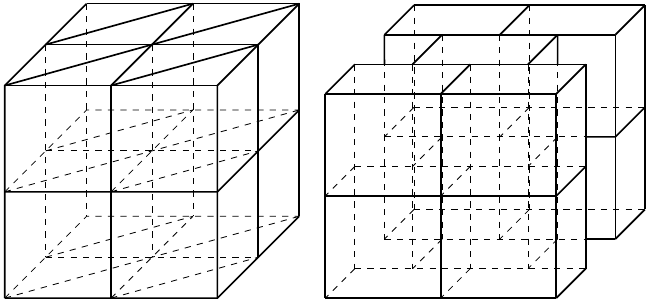}
\caption{\small Deformed triangular and honeycomb lattices in 3D}\label{3d tri}
\end{center}
\end{figure}

The methodology used to define 3D algebraic area can be extended to other 3D lattices, such as deformed triangular and honeycomb lattices (see figure \ref{3d tri}). However, the associated enumeration formulae and their connection with quantum exclusion statistics remain unresolved issues that require further study. Additionally, exploring the algebraic area enumeration for {\it open} random walks on various 3D lattices would also be of interest (see \cite{Desbois, open} for open walks on a square lattice).

\noindent \textbf{Acknowledgment} The author would like to express gratitude to St\'ephane Ouvry and Alexios P. Polychronakos for valuable discussions. The author also acknowledges the financial support of China Scholarship Council (No. 202009110129).

\section*{Appendix A: Recursive relation for enumeration of cubic lattice walks}

Consider an ${\bf n}$-step cubic lattice walk that consists of $m_1$ steps in the direction $(1,0,0)$, $m_2$ steps in the direction $(-1,0,0)$, $l_1$ steps in the direction $(0,1,0)$, $l_2$ steps in the direction $(0,-1,0)$, $r_1$ steps in the direction $(0,0,1)$, $r_2$ steps in the direction $(0,0,-1)$, where $m_1+m_2+l_1+l_2+r_1+r_2={\bf n}$.
If the walk is open, we can close it by adding a straight line that connects the endpoint to the starting point.
Let $C_{m_1,m_2,l_1,l_2,r_1,r_2}(A)$ denote the number of such walks that enclose a 3D algebraic area $A$. The generating function $Z_{m_1,m_2,l_1,l_2,r_1,r_2}(\Q)=\sum_{A}C_{m_1,m_2,l_1,l_2,r_1,r_2}(A)\Q^A$ can be computed by the recursion
{\small \bea \nonumber
Z_{m_1,m_2,l_1,l_2,r_1,r_2}(\Q)
&=&
\Q^{(l_2-l_1+r_1-r_2)/2} Z_{m_1-1,m_2,l_1,l_2,r_1,r_2}(\Q) +
\Q^{(l_1-l_2+r_2-r_1)/2} Z_{m_1,m_2-1,l_1,l_2,r_1,r_2}(\Q) \\\nonumber & & +
\Q^{(m_1-m_2+r_2-r_1)/2} Z_{m_1,m_2,l_1-1,l_2,r_1,r_2}(\Q)+
\Q^{(m_2-m_1+r_1-r_2)/2} Z_{m_1,m_2,l_1,l_2-1,r_1,r_2}(\Q) \\\nonumber & & +
\Q^{(l_1-l_2+m_2-m_1)/2} Z_{m_1,m_2,l_1,l_2,r_1-1,r_2}(\Q)+
\Q^{(l_2-l_1+m_1-m_2)/2} Z_{m_1,m_2,l_1,l_2,r_1,r_2-1}(\Q)\\\label{recursion}
\eea}
with $Z_{0,0,0,0,0,0}(\Q)=1$ and $Z_{m_1,m_2,l_1,l_2,r_1,r_2}(\Q)=0$ whenever $\min(m_1,m_2,l_1,l_2,r_1,r_2)<0$.

For closed walks of length ${\bf n}=2n$, we have
\be \label{recursion bis}
\sum_{A}C_{2n}(A)\Q^A=\sum_{m=0}^{n}\sum_{l=0}^{n-m} Z_{m,m,l,l,n-m-l,n-m-l}(\Q).
\ee

Table \ref{table} lists the first few values of $C_{2n}(A)$ obtained from either the recursion (\ref{recursion}) and (\ref{recursion bis}), or the general term formula (\ref{final formula}).

\begin{table}[H]
\begin{center}
\begin{tabular}{rrrrrr}
\hline
\multicolumn{1}{c}{} & $2n=2$ & 4 & 6 & 8 & 10 \\ \hline
$A=~~0$ & 6 & 66 & 948 & 16626 & 338616 \\
$\pm 1$ & & 24 & 756 & 19392 & 483420 \\
$\pm 2$ & & & 144 & 6744 & 230340 \\
$\pm 3$ & & & 12 & 1584 & 82980 \\
$\pm 4$ & & & & 336 & 27000 \\
$\pm 5$ & & & & 48 & 7740 \\
$\pm 6$ & & & & & 1980 \\
$\pm 7$ & & & & & 420 \\
$\pm 8$ & & & & & 60 \\
\multicolumn{1}{c}{Total counting} & 6 & 90 & 1860 & 44730 & 1172556 \\ \hline
\end{tabular}
\end{center}
\caption{$C_{2n}(A)$ up to $2n=10$ for cubic lattice walks of length $2n$.}\label{table}
\end{table}

\section*{Appendix B: Combinatorial results for $(1,1,2)$-compositions and $c_{1,1,2}$}

\noindent 1. By considering the combinatorial interpretation of cluster coefficient $b'_n$ as fermions of two types and two-fermion bound states, we can derive the counting of $(1,1,2)$-compositions of $2n$ with $\tilde{l}_1+\ldots+\tilde{l}_{j+1}=\tilde{l}'_1+\ldots+\tilde{l}'_{j+1}$ to be
\be \nonumber
N_{1,1,2}(n)=1+\sum_{k=0}^{n-1}\sum_{m=0}^{k}\binom{k}{m}  \binom{n+m-k}{m+1}^2=1, 2, 7, 27, 108, 443, \ldots
\ee
with the convention $N_{1,1,2}(0)=1$.
Equivalently, the generating function of the $N_{1,1,2}(n)$'s is
\be \nonumber
\sum_{n=0}^{\infty}x^n N_{1,1,2}(n)=\frac{1-x}{\sqrt{x^4-2 x^3+7 x^2-6 x+1}}.
\ee

\noindent 2. We have
$$2n\sum_{\substack{\tilde{l}_1,\ldots,\tilde{l}_{j+1};\tilde{l}'_1,\ldots,\tilde{l}'_{j+1};l_1,\ldots,l_j \\ (1,1,2){\text{-composition of }}2n\\ \tilde{l}_1+\ldots+\tilde{l}_{j+1}=\tilde{l}'_1+\ldots+\tilde{l}'_{j+1}\\l_1+\ldots+l_j=k}}
c_{1,1,2}(\tilde{l}_1,\ldots,\tilde{l}_{j+1}; \tilde{l}'_1,\ldots,\tilde{l}'_{j+1}; l_1,\ldots,l_j)=\binom{2n}{n} \binom{n}{k}^2,$$
\noindent from which we infer
$$2n
{\hspace{-0.5em}} \sum_{\substack{\tilde{l}_1,\ldots,\tilde{l}_{j+1};\tilde{l}'_1,\ldots,\tilde{l}'_{j+1};l_1,\ldots,l_j \\ (1,1,2){\text{-composition of }}2n\\ \tilde{l}_1+\ldots+\tilde{l}_{j+1}=\tilde{l}'_1+\ldots+\tilde{l}'_{j+1}}} {\hspace{-0.5em}}
c_{1,1,2}(\tilde{l}_1,\ldots,\tilde{l}_{j+1}; \tilde{l}'_1,\ldots,\tilde{l}'_{j+1}; l_1,\ldots,l_j)
=\sum_{k=0}^{n}\binom{2n}{n} \binom{n}{k}^2
=\binom{2n}{n}^2.$$

\noindent 3.
\noindent In the limit $q\to \infty$ \cite{Shuang,papers}
$$\frac{1}{q}\sum_{k=1}^{q-j}
t_k^{\tilde{l}_1}\bar{t}_k^{\tilde{l}'_1}s_k^{l_1}
t_{k+1}^{\tilde{l}_2}\bar{t}_{k+1}^{\tilde{l}'_2}s_{k+1}^{l_2}\ldots\to
\binom{2(l_1+\dots+l_j)}{l_1+\ldots+l_j},
$$
we recover the overall counting to be
\bea \nonumber & &
2n
{\hspace{-0.5em}} \sum_{\substack{\tilde{l}_1,\ldots,\tilde{l}_{j+1};\tilde{l}'_1,\ldots,\tilde{l}'_{j+1};l_1,\ldots,l_j \\ (1,1,2){\text{-composition of }}2n\\ \tilde{l}_1+\ldots+\tilde{l}_{j+1}=\tilde{l}'_1+\ldots+\tilde{l}'_{j+1}}} {\hspace{-0.5em}}
c_{1,1,2}(\tilde{l}_1,\ldots,\tilde{l}_{j+1}; \tilde{l}'_1,\ldots,\tilde{l}'_{j+1}; l_1,\ldots,l_j)
\binom{2(l_1+\dots+l_j)}{l_1+\ldots+l_j}\\\nonumber
&=&
\sum_{k=0}^{n}
\sum_{\substack{\tilde{l}_1,\ldots,\tilde{l}_{j+1};\tilde{l}'_1,\ldots,\tilde{l}'_{j+1};l_1,\ldots,l_j \\ (1,1,2){\text{-composition of }}2n\\ \tilde{l}_1+\ldots+\tilde{l}_{j+1}=\tilde{l}'_1+\ldots+\tilde{l}'_{j+1}\\l_1+\ldots+l_j=k}}
2n \, c_{1,1,2}(\tilde{l}_1,\ldots,\tilde{l}_{j+1}; \tilde{l}'_1,\ldots,\tilde{l}'_{j+1}; l_1,\ldots,l_j)
\binom{2k}{k} \\\nonumber
&=& \binom{2n}{n}\sum_{k=0}^{n}
\binom{n}{k}^2  \binom{2k}{k},
\eea
which is indeed $[x^0 y^0 z^0](x+y+z+x^{-1}+y^{-1}+z^{-1})^{2n}
= 6, 90, 1860, 44730, 1172556,\ldots$ (see the \href{https://oeis.org/A002896}{OEIS sequence A002896}).

\end{document}